\documentclass[%
reprint,
superscriptaddress,
 amsmath,amssymb,
 aps,prl,twocolumn
]{revtex4-1}

\usepackage[utf8]{inputenc}
\usepackage[english]{babel}
\usepackage{graphicx}% Include figure files
\usepackage{dcolumn}% Align table columns on decimal point
\usepackage{bm}% bold math
\usepackage{braket}
\usepackage{color}
\usepackage{enumerate}
\usepackage{hyperref}% add hypertext capabilities

\usepackage{hyperref}
\hypersetup{pdfborder=0 0 0,colorlinks=true,citecolor=blue,linkcolor=blue}

\allowdisplaybreaks

\begin{document}

\title{Phase transitions of hybrid perovskites simulated by machine-learning force fields trained on-the-fly with Bayesian inference}
\author{Ryosuke Jinnouchi}
\affiliation{University of Vienna, Faculty of Physics and Center for Computational Materials
Sciences, Sensengasse 8/12, 1090 Vienna}
\affiliation{Toyota Central R\&{}D Labs, Inc., 41-1, Yokomichi, Nagakute, Aichi 480-1192, Japan}
\author{Jonathan Lahnsteiner}
\affiliation{University of Vienna, Faculty of Physics and Center for Computational Materials
Sciences, Sensengasse 8/12, 1090 Vienna}
\author{Ferenc Karsai}
\affiliation{VASP Software GmbH, Sensengasse 8, 1090 Vienna, Austria}
\author{Georg Kresse}
\author{Menno Bokdam}
\email{menno.bokdam@univie.ac.at}
\affiliation{University of Vienna, Faculty of Physics and Center for Computational Materials
Sciences, Sensengasse 8/12, 1090 Vienna}

\date{\today}% It is always \toda, today,

\begin{abstract}
Realistic finite temperature simulations of matter are a formidable challenge for first principles methods. Long simulation times and large length scales are required, demanding years of compute time. Here we present an on-the-fly machine learning scheme that generates force fields automatically during molecular dynamics simulations. This opens up the required time and length scales, while retaining the distinctive chemical precision of first principles methods and minimizing the need for human intervention. The method is widely applicable to multi-element complex systems. We demonstrate its predictive power on the entropy driven phase transitions of hybrid perovskites, which have never been accurately described in simulations. Using machine learned potentials, isothermal-isobaric simulations give direct insight into the underlying microscopic mechanisms. Finally, we relate the phase transition temperatures of different perovskites to the radii of the involved species, and we determine the order of the transitions in Landau theory.
\end{abstract}

\maketitle

Predicting the finite temperature properties of materials from first principles (FP) has always been a dream of materials scientists but it has hardly been achieved except for the simplest of solids. The main obstacle is that the required system sizes and simulation times are simply not attainable using standard FP techniques. Training force fields using machine learning (ML) techniques is an obvious solution to the problem. However, what has prevented ML from being widely applied is the construction of suitable reference structures. In conventional approaches\cite{Behler:prl07,Bartok:prl10}, training structures are selected using chemical intuition, FP calculations are performed for them, and machine-learned force fields (MLFF) are fitted. Later, when the user realizes that structures outside the present training set need to be included, additional structures are added, and the force field is retrained. This is a time-consuming trial and error process often taking months for a single material, and it is practically untraceable for multi-elemental complex materials. On-the-fly machine learning has been suggested as an alternative possibly reducing human intervention\cite{Li:prl15}. The prime progress in the present work is that the predicted (Bayesian) error is used to decide whether FP calculations are required or can be bypassed. We put our generally applicable algorithm to the test by applying it to a puzzling material exhibiting very fast hydrogen dynamics, as well as very slow rotational dynamics. The on-the-fly ML allows us to predict phase diagrams with FP quality so far unprecedented efficiency.

We have chosen hybrid perovskites as a first application of our scheme, because the slow rotational dynamics of the molecules makes straightforward FP molecular dynamics exceedingly time-consuming. Furthermore, hybrid perovskites possess a huge scientific and technological potential. Methylammonium (MA)PbI$_3$ is a promising solar cell material\cite{Hirasawa:jpsj94,Kojima:jacs17,solchart:18} with a high charge-carrier mobility\cite{Stranks:sc13}. Many experimental and theoretical studies have been performed on its atomic structure and dynamical properties\cite{Weber:zfn78,Onoda-Yamamuro:jpcs90,Baikie:jmca:13,Kawamura:jpsj02,Stoumpos:ic13,Whitfield:sr16,Poglitsch:jcp87,Wasylishen:ssc85,Chen:pccp15,Filippetti:prb14,Mattoni:jpcc15,Lahnsteiner:prb16,Bokdam:prl17,Lahnsteiner:prm18} and they have revealed that experimentally this material exhibits two entropy-driven phase transitions from an orthorhombic to a tetragonal phase at 160~K, and from a tetragonal to a cubic phase at 330~K. Estimates of the transition temperatures from FP have not been reported to date and are elusive to be obtained using standard FP techniques alone, owing to the fact that the transitions are entropy driven. Although in the orthorhombic phase the molecules and the cage are essentially frozen, in the cubic phase the MA molecules and the cage reorient rapidly exploring a large phase space. Moreover, recent theoretical studies\cite{Lahnsteiner:prm18} indicate that the available semiempirical force fields are not accurate and that only few fairly expensive density functionals describe the instabilities of the cage and the interaction between the molecules and the cage with sufficient accuracy\cite{Bokdam:prl17}.
%-------------------------------------------------------------------------
\begin{figure}
    \begin{center}
    \includegraphics[width=.88\columnwidth ,clip=true]{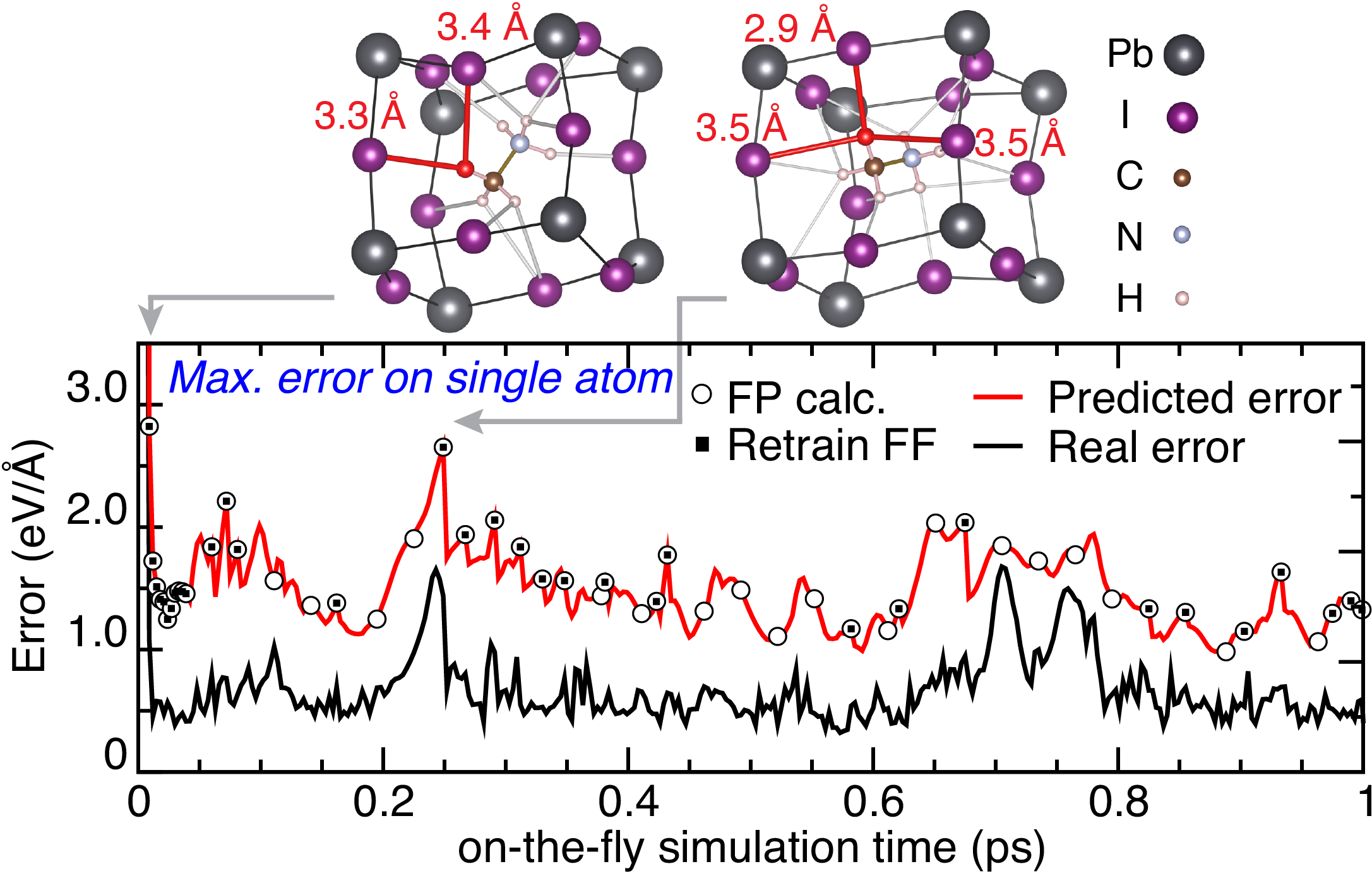}
    \end{center}
   \caption{The error in the force field during the first pico-second of the on-the-fly simulation. The predicted (Bayesian) error for the unitless force (provided by Eq.~(\ref{eq_Bayes})) closely resembles the real error. A part of the structure at 0 and 0.24 ps is shown on top. The hydrogen atom that exhibits the largest real and predicted errors at 0.24 ps is drawn as a red sphere, and the H-I bonds shorter than 3.6 $\mathrm{\AA}$ as red/gray lines. The simulation is executed on MAPbI$_3$ at 450 K.}
\label{fig:1}
\end{figure}
%-------------------------------------------------------------------------

For the description of the machine learned potential energy surface, we use a variant of the Gaussian approximation potential (GAP) pioneered by Bar\'tok and coworkers\cite{Bartok:prl10}. In GAP, the potential energy $U$ of a system with $N_{\mathrm{a}}$ atoms is described as a summation of {\em local} atomic potential energies $U_i$,
\begin{equation}
 U=\sum\limits_{i=1}^{N_\mathrm{a}} U_{i} = \sum\limits_{i=1}^{N_\mathrm{a}}\sum\limits_{i_{\mathrm{B}}=1}^{N_\mathrm{B}} w_{i_{\mathrm{B}}} K\left(\mathbf{X}_{i},\mathbf{X}_{i_{\mathrm{B}}}\right). 
 \label{eq_apes}
\end{equation}
Each $U_{i}$ is expressed as a linear combination of the kernel function $K\left(\mathbf{X}_{i},\mathbf{X}_{i_{\mathrm{B}}}\right)$ and weight factors $w_{i_{\mathrm{B}}}$. The kernel measures the similarity between the local configuration around atom $i$ and the reference local configuration $i_{\mathrm{B}}$. For the descriptor $\mathbf{X}_{i}$ and the kernel $K$, we adopted a variant of the Smooth Overlap Atomic Positions (SOAP)\cite{Bartok:prb13} (see Supplemental Materials (SM)\cite{SM}). The Eq.~(\ref{eq_apes}) allows to describe the energy, forces and stress tensor (EFS) for a given structure as $\bm{\phi} \mathbf{w}$. Here, $\mathbf{w}=\{w_{i_{\mathrm{B}}}\}$ and $\bm{\phi}$ is a matrix containing $K\left(\mathbf{X}_{i},\mathbf{X}_{i_{\mathrm{B}}}\right)$ and its derivatives with respect to the coordinates and lattice vectors. Similarly, the EFSs on all training structures can be summarized as $\bm{\Phi} \mathbf{w}$, where $\bm{\Phi}$ collects $\bm{\phi}$ for all training structures. The Bayesian theorem~\cite{Bishop:book06} allows us to determine $\mathbf{w}$ and the uncertainty $\bm{\sigma}$ in the predicted EFS as
\begin{align}
\mathbf{w} &= \left[ \mathbf{I}/ \sigma_{\mathrm{w}}^{2} + \mathbf{\Phi}^{\mathrm{T}} \mathbf{\Phi} / \sigma_{\mathrm{v}}^{2} \right]^{-1} \mathbf{\Phi}^{\mathrm{T}}\mathbf{T} / \sigma_{\mathrm{v}}^{2},
\label{equation_wm}
\end{align}
\begin{equation}
 \bm{\sigma}=\sigma_{\mathrm{v}}^{2}\mathbf{I}+\bm{\phi}^{\mathrm{T}}\left[\mathbf{I}/\sigma_{\mathrm{w}}^{2}+\mathbf{\Phi}^{\mathrm{T}}\mathbf{\Phi}/\sigma_{\mathrm{v}}^{2}\right]^{-1}\bm{\phi}.
 \label{eq_Bayes}
\end{equation}
The FP data of the training structures enter in the vector $\mathbf{T}$. $\mathbf{I}$ is the identity matrix. The parameters $\sigma_{\mathrm{v}}^{2}$ and $\sigma_{\mathrm{w}}^{2}$ are determined to balance the accuracy and robustness of the MLFF using the evidence approximation~\cite{Bishop:book06}. The on-the-fly scheme has been integrated within the VASP code~\cite{Kresse:prb96,Kresse:cms96} (section A in SM\cite{SM}).

The actual training was performed using a state-of-the-art meta-gradient corrected functional\cite{Sun:prl15} and running extensive FP simulations in all three experimentally known phases using $2\times2\times2$ unit cells\cite{SM}. To determine whether FP calculations are required, the Bayesian error Eq.~(\ref{eq_Bayes}) is used (see Fig.~\ref{fig:1}). If the error is above a certain threshold, adjusted also on the fly, FP calculations are performed as indicated by the white circles. The FP data are used to refine the force field at the MD steps shown as black dots. In this manner, if the system stays in a local minimum, most FP calculations are bypassed, but after reordering of  the PbI cage  or the MA molecules FP calculations are performed as illustrated in Fig.~\ref{fig:1}. 
The efficiency of the on-the-fly learning is demonstrated by the fact that during the training  99~\% of the FP calculations are skipped reducing the computational time by almost a factor of 100 even during learning. This enables us to extensively explore the phase space involving slow molecular reorientations in the PbI$_3$ framework occurring on a ps time scale. The generated regression model predicts energies, forces and stress tensors with near-FP quality of 2.6~meV/atom, 0.07~eV/\AA{} and 0.82~kbar, respectively (Section C in SM\cite{SM}). 
%-------------------------------------------------------------------------
\begin{figure*}
    \begin{center}
    \includegraphics[width=1.9\columnwidth ,clip=true]{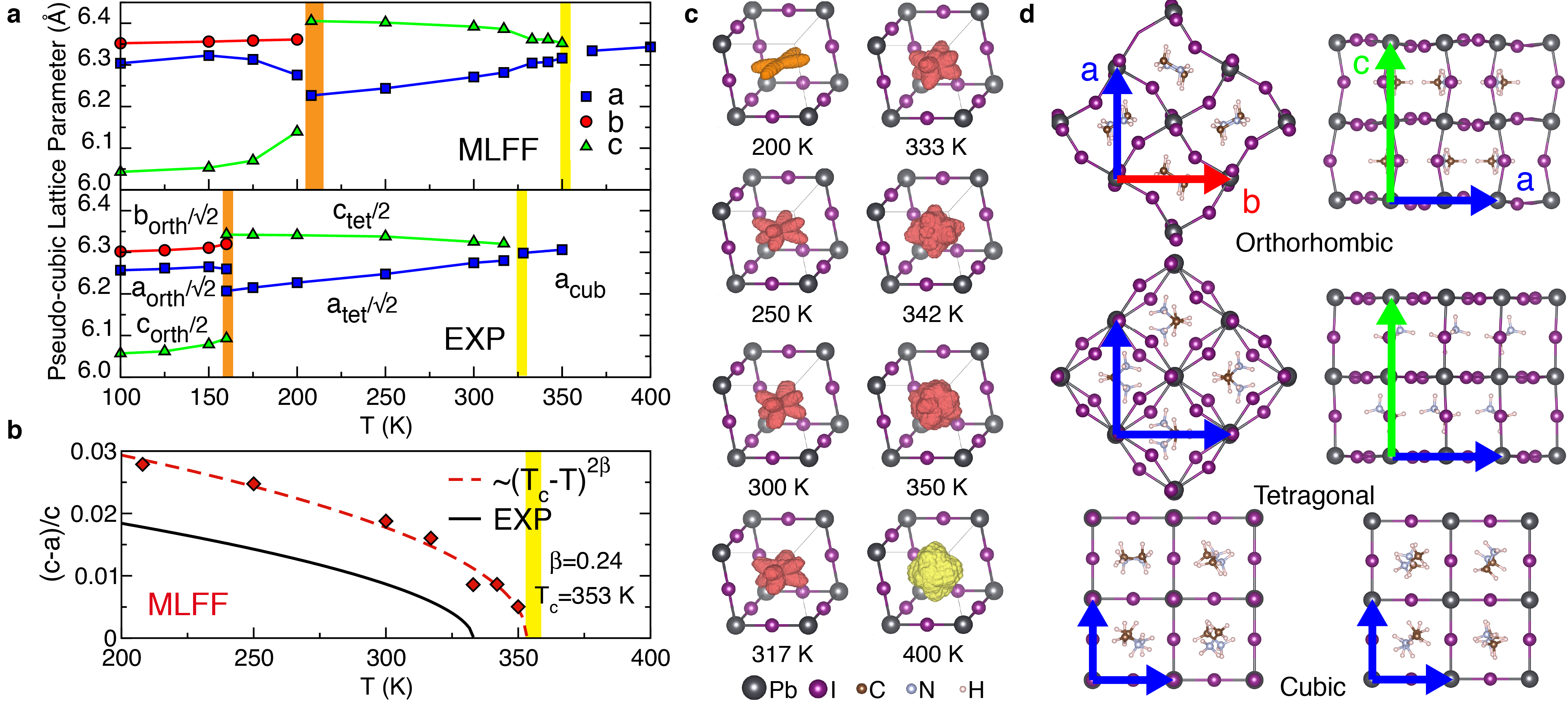}
    \end{center}
   \caption{Phase transitions of MAPbI$_3$. (\textbf{a}) Simulated lattice constants compared to experiment. (\textbf{b}) A power law, $(T_c-T)^{2\beta}$, fitted to the simulated and experimental tetragonal distortion, $(c-a)/c$, where $T_c$ and $\beta$ are the transition temperature and the critical exponent, respectively. (\textbf{c}) 3-dimensional polar plots of the probability distributions of the MA molecules (C-N bond orientation) at various temperatures, ranging from 200~K to 400~K. (\textbf{d}) Schematic representation of the three MAPbI$_3$ phases as obtained by the MLFF. Experimental data shown in (\textbf{a}) and (\textbf{b}) are taken from Ref.~\cite{Whitfield:sr16}, and vertical orange and yellow bars indicate the orthorhombic to tetragonal and the tetragonal to cubic phase transition temperatures, respectively.}
\label{fig:2}
\end{figure*}
%-------------------------------------------------------------------------

In Figure~\ref{fig:2}(a), we show the simulated lattice constants of MAPbI$_3$ as a function of the temperature and compare them with experiment\cite{Whitfield:sr16}. For this simulation $4\times4\times4$ unit cells were used (for determination of lattice constants see section D in SM\cite{SM}). The constructed force field accurately reproduces the structure of all three phases. An analysis on the tetragonal distortion, $(c-a)/c$, shown in Fig.~\ref{fig:2}(b) allows to pin-point the tetragonal to cubic phase transition temperature at 353~K. The critical exponent of 0.24 agrees well with the reported experimental results of 0.22-0.285\cite{Whitfield:sr16} and a theoretically expected value of $1/4$ for a tri-critical point on the basis of Landau theory. A careful free energy “umbrella sampling”\cite{Kastner:cmos11} analysis (details in the SM\cite{SM}) was used to determine the transition temperature of 215$\pm$10~K between the orthorhombic and tetragonal phase. Furthermore, for the orthorhombic to tetragonal transition, the change of the entropy at this phase transition is 1.3$\pm$0.6~$k_{\rm B}$ per MA molecule agreeing reasonably with the experimental value of 2.3~$k_{\rm B}$\cite{Onoda-Yamamuro:jpcs90}. The theoretical results compare well with the available experimental data and are in essence only limited by the accuracy of the density functional.

Contrary to experiment, our simulations readily provide atomic-scale insight into the entropy-driven phase transitions. As a first step, we have analyzed the orientation of the molecular C-N axis. Figure~\ref{fig:2}(c) shows 3-dimensional polar plots of the probability distribution of the molecular orientation in the PbI$_3$ framework at 200-400~K. In the orthorhombic phase at 200~K, the polar distribution exhibits two specific orientations predominantly along the $x$- or $y$-axis. In this phase the molecules are frozen, and their orientation alternates only spatially. In the tetragonal phase at 250~K, the molecules are also canting in $+z$ and $-z$ direction out of the $xy$-plane, so that eight lobes are visible. It should be noted that our previous FP MD simulations did not describe the molecular order in the tetragonal phase accurately at lower temperatures either because of the short simulation time or the fixed volume\cite{Bokdam:prl17}. In the present simulations, at 250~K the short range molecular order is consistent with the order shown in the snapshot of Fig.~\ref{fig:2}(d) for the tetragonal phase. Between 300 and 350~K a spherical probability distribution gradually develops, indicating that the molecules continuously obtain additional rotational freedom close to the tetragonal to cubic phase transition. In the cubic phase at 400~K, the molecules realize a nearly free rotation, whereas in the tetragonal phase, the molecules exhibit hindered reorientations. Specifically, at 350~K we predict a reorientation rate of 6.4~ps (Section F in the SM~\cite{SM}) agreeing well with the experimental results of 1.0-5.4~ps at 300-350~K\cite{Onoda-Yamamuro:jpcs90,Poglitsch:jcp87,Chen:pccp15}.

\begin{figure*}
    \begin{center}
    \includegraphics[width=2.075\columnwidth ,clip=true]{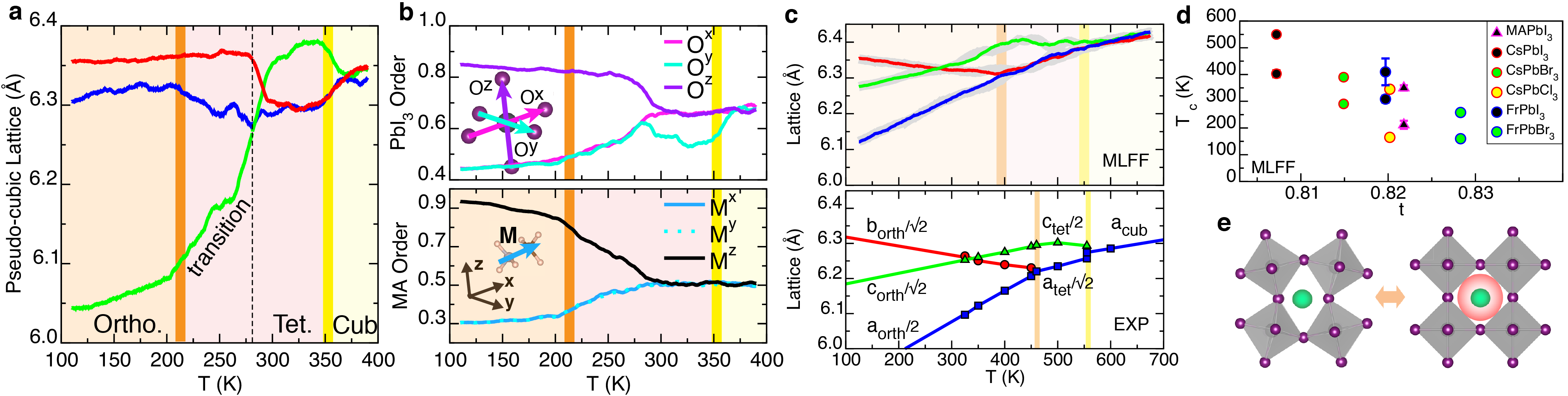} 
    \end{center}
\caption{Dynamics of MAPbI$_3$ and CsPbI$_3$ lattice upon heating/cooling MDs at a rate of 0.5~K/ps. Lattice constants (\textbf{a}) and order parameters (\textbf{b}) provided by the heating simulation for a $4\times4\times4$ unit cell of MAPbI$_3$. The black dashed line in a indicates a switch from the orthorhombic to tetragonal representation. The bond vectors used in the order parameters are sketched in the inset. (\textbf{c}) Simulated lattice constants for a $6\times6\times6$ unit cell of CsPbI$_3$ compared with experiments of CsPbI$_3$. Results for heating are shown using colored lines, whereas results for cooling are shown using gray lines. Experimental data shown in (\textbf{c}) are taken from Ref.~\cite{Marronnier:acsn18}. (\textbf{d}) Phase transition temperatures of perovskites in dependence on the Goldschmidt tolerance factor. The circles and triangles represent the data for various $ABX_3$ perovskites. Lattice and order parameters for these perovskites are presented in the SM\cite{SM}. (\textbf{e}) Sketch of the thermal radius ($r_A$) of the cation.}
\label{fig:3}
\end{figure*}

To obtain more insight on the atomic scale mechanism of the phase transitions, we performed simulations on $4\times4\times4$ unit cells using a slow heating rate of 0.5~K/ps. Even with state-of-the-art massively-parallel computers such a simulation would take several years using FP techniques alone. To unravel the microscopic mechanism of the phase transition, we introduce order parameters denoted as $\mathbf{O}$ (octahedron) and $\mathbf{M}$ (molecular) as sketched in Fig.~\ref{fig:3}(b). The vector $\mathbf{O}$ measures the angular correlation between adjacent PbI$_6$ octahedra along the $x$-, $y$- and $z$-axis and resolves the ordering of the frame. $O^x$, for example, approaches unity when adjacent octahedra along the $x$-axis are tilted in the same direction, while it approaches zero when adjacent octahedra are tilted oppositely. Similarly, the vector $\mathbf{M}$ measures the angular correlation between adjacent MA molecules and resolves the molecular order\cite{SM}. Figure~\ref{fig:3}~(b) shows that below 220~K adjacent octahedra are tilted in the same direction along the $z$-axis, while they are oppositely rotated in the $xy$-plane. A corresponding trend appears in the molecular order parameter. This is the typical pattern of the orthorhombic phase as illustrated in Fig.~\ref{fig:2}(d). Between 220~K and 270~K, thermal fluctuations allow the molecules to reorient, and the three elements of the molecular order parameter gradually merge to 0.5, which corresponds to the 90-degree angle between neighboring molecules in the tetragonal phase. The octahedral order parameter changes rather abruptly to the tetragonal pattern after the transition in the molecular order parameter has finished (drop of $O^y$ in Fig.~\ref{fig:3}(b) at 270~K).  These results clearly indicate that the transition of the PbI$_6$ octahedra occurs after the reorganization of the molecules has completed, in other words, the molecules seem to inhibit the transition. The exact same trend is observed in a constant temperature MD at 220 K shown in Section D in the SM\cite{SM}.

All in all, our simulations suggest that the orthorhombic to tetragonal transition is first order. The transition region in Figs.~\ref{fig:3}(a) and (b) is a result of the still fairly fast heating compared to experiment and slow dynamics of the MA molecules. The second transition from the tetragonal to the cubic phase, however, is continuous, since the lattice parameters and order parameters evolve smoothly. This also agrees with the reversibility, i.e. upon cooling the cubic to tetragonal transition is readily observed (see Fig. D1 in the SM\cite{SM}) and the hysteresis between heating and cooling is small.

%-------------------------------------------------------------------------
%-------------------------------------------------------------------------

The phase diagram of MAPbI$_3$ is well known experimentally. In order to test the methodology for a less well studied material, we move to CsPbI$_3$, where the experimentally observed phases strongly depend on the process of heating and annealing as well as sample preparation, and thus, their interpretations are still controversial\cite{Marronnier:acsn18,Sutton:acsel18}. Figure~\ref{fig:3}(c) shows the simulated lattice constants as a function of the temperature. The force-field generation and MD simulations are described in the SM\cite{SM}. Our MD simulations show two now clearly continuous phase transitions between orthorhombic and tetragonal at 403$\pm$13~K, and between tetragonal and cubic at 550$\pm$5~K. These transitions are also fully reversible and are observed both under heating (shown as colored lines) and cooling (shown as gray lines). The simulated lattice constants agree well with the experimental results reported by Marronier~\textit{et al.}\cite{Marronnier:acsn18} The difference between MAPbI$_3$ and CsPbI$_3$ is related to the slow, extra rotational degrees of freedom of the MA molecule, which become accessible only in the tetragonal and cubic phases, whereas for Cs only lateral “rattling” movements become progressively more pronounced upon heating.

Although both perovskites exhibit similar crystallographic structures, the transition temperatures of MAPbI$_3$ are significantly lower than those of CsPbI$_3$. In order to better understand the origin of this difference, we further extend our on-the-fly scheme to other inorganic perovskites $ABX_3$ ($A$=Cs or Fr, $B$=Pb, and $X$=I, Br or Cl). Comparison of the transition temperatures with the Goldschmidt tolerance factor\cite{Goldschmidt:27}, $t=(r_A+r_X)/(\sqrt{2}(r_{\rm Pb}+r_{X}))$, indicates that the different transition temperatures relate reasonably well with the size of the ions. Here, $r_{A}$, $r_{\rm Pb}$ and $r_{X}$ denote the ionic radii of cation $A$, Pb and halogen $X$, respectively. The determination of these values is described in Section G in the SM\cite{SM}. Figure~\ref{fig:3}(d) illustrates that the transition temperatures of the hybrid perovskites decrease with increasing tolerance factor. This trend can be explained by the mechanism described below. The calculated tolerance factors of the inorganic perovskites indicate that the PbX$_6$ octahedra need to tilt in order to allow the formation of multiple bonds between the cation $A$ and the halogen $X$. Without tilting, the Pb-halogen distance ($r_{\rm Pb}+r_X$) is too large compared to the optimal cation-halide distance ($r_A+r_X$). The prediction is consistent with the fact that all six perovskites exhibit the orthorhombic phase at low temperature. The situation, however, changes with rising temperature. The effective radius of the cation $A$ increases by thermal fluctuations (as schematically shown in Fig.~\ref{fig:3}(e)) and the tilting of the octahedra becomes unnecessary. Thus, the phase transition occurs. The necessary amount of the thermal fluctuation depends on the radii. The larger the cation $A$ or the smaller the halide $X$ radius, the smaller is the necessary fluctuation. Therefore, the transition temperatures decrease with increasing tolerance factor.\newline

{\em In summary}, we have shown here that the combination of FP calculations with on-the-fly ML has the potential to lead to a paradigm shift in the modelling of complex materials at finite temperature. On-the-fly ML enables an exceedingly efficient sampling of structures over a large phase space with very little human intervention. Our scheme is straightforwardly applicable to complex multi-elemental materials and can be easily applied to other materials science problems such as ionic diffusion or catalytic reactions. For hybrid perovskites, we obtain excellent qualitative agreement with experiment as well as useful new insights. We observe the temperature driven transition from the orthorhombic, over the tetragonal to the cubic phase  to be omnipresent in Pb-chalcogenides. However, MAPbI$_3$ is in fact unique  with a first-order orthorhombic to tetragonal phase transition, related to the unfreezing of the molecular rotational degrees of freedom.

\begin{acknowledgments}
J.L. and M.B. gratefully acknowledge funding by the Austrian Science Fund (FWF): P 30316-N27. Computations were partly performed on the Vienna Scientific Cluster VSC3. All authors gratefully thank Ryoji Asahi for many suggestions on applications of the machine-learning method to materials science and Carla Verdi for proof reading of the manuscript.
\end{acknowledgments}

\end{document}